# A high dynamic Micro Strips Ionization Chamber featuring Embedded Multi DSP Processing

Francesco Voltolina, Ralf H. Menk and Sergio Carrato

On behalf of the SAXS Detector Collaboration

***Abstract*--An X-ray detector will be presented that is the combination of a segmented ionization chamber featuring one-dimensional spatial resolution integrated with an intelligent ADC front-end, multi DSP processing and embedded PC platform. This detector is optimized to fan beam geometry with an active area of 192 mm (horizontal) and a vertical acceptance of 6 mm. Spatial resolution is obtained by subdividing the anode into readout strips, having pitch of 150 micrometers, which are connected to 20 custom made integrating VLSI chips (each capable of 64-channel read-out and multiplexing) and read out by 14 bits 10 MHz ADCs and fast adaptive PGAs into DSP boards.**

**A bandwidth reaching 3.2Gbit/s of raw data, generated from the real time sampling of the 1280 micro strips, is cascaded processed with FPGA and DSP to allow data compression resulting in several days of uninterrupted acquisition capability.**

**Fast acquisition rates reaching 10 kHz are allowed due to the MicroCAT structure utilized not only as a shielding grid in ionization chamber mode but also to provide active electron amplification in the gas.**

## I. Introduction

This detector is primarily aimed to SAXS experiments (see [1], [2]) and appears, from the user point of view, like a "black box": it can interpret some commands, transferred generally via the local area network connection, and consequentially performs some actions leading to a result, in the form of a file, which is transferred back to the user. More in detail (see Fig. 1) we can identify a chain of *three main blocks* linking the USER to the JAMEX ASICs (that are directly responsible for the physical readout of the 1280 microstrips, see [3]):

- **PC Embedded** hosting the Sundance system and implementing the USER interface
- **Sundance Acquisition System** responsible for the Real Time signal acquisition and processing
- **J2S Board** offering programmable analog signal conditioning and digital signal routing

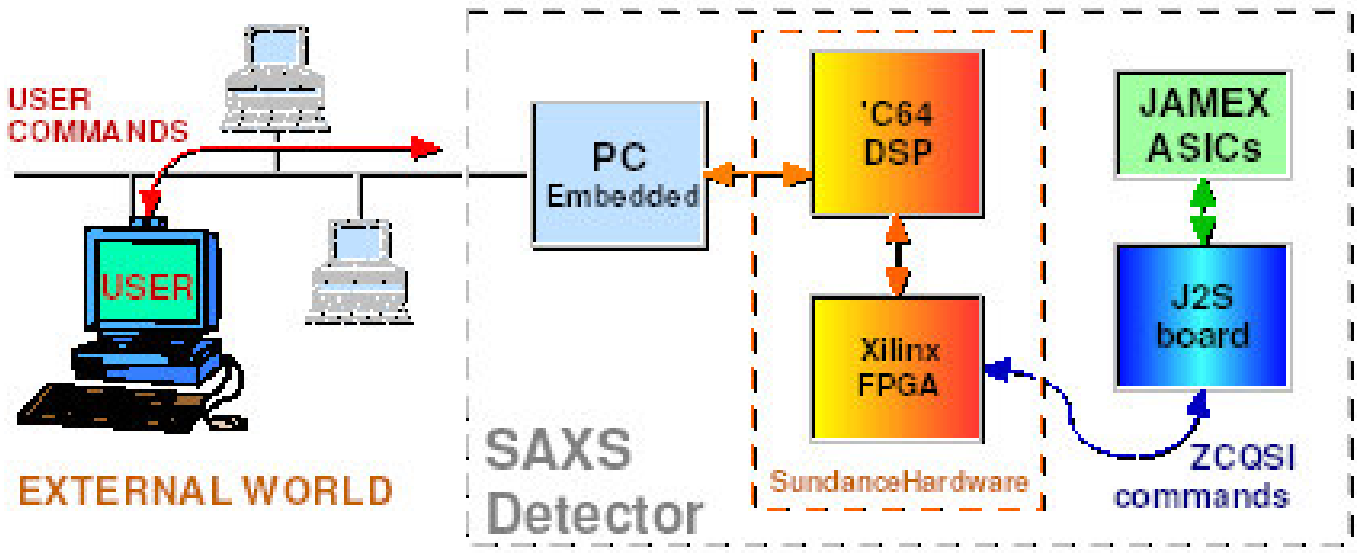

Fig. 1. Main blocks and command paths in the SAXS detector.

These three blocks interact with each other by means of exchanging specific commands via dedicated interfaces: the PC is communicating to the Sundance system via the PCI bus, while the Sundance System controls the J2S Board via an interface and a set of command that we have called `ZCQSI` (Zero Clock Quiet Serial Interface, indicating the aim to specify a low noise interface method using a reduced number of wires). We must mention the important commands and data path connecting the J2S board to the 10 JAMEX Hybrids: the characteristics of this interfacing will be explained in Chapter III that is dedicated to the detector hardware.

## II. Functional description of the detector

A consistent set of *user commands* is the base for the configuration and development (involving both software and hardware design) of all the previously mentioned main blocks. In order to familiarize with some technical terms used in the definition of this command set it is necessary to give some more information on how this SAXS detector works: result of the *measurement* is an `Image` (a bidimensional matrix stored in a file) for which we chose a maximum size of 1280 by 2048 words. The DSP used in the Sundance System is a 32-bit

Manuscript received October 31, 2004. The author is individually supported by the European Community – Research Infrastructure Action under the FP6 "Structuring the European Research Area" Programme (through the Integrated Infrastructure Initiative "Integrating Activity on Synchrotron and Free Electron Laser Science").

The research presented in this article has been additionally supported by the European Community contract no. FMBICT980104.

Francesco Voltolina is with the Institut für Höchstfrequenztechnik und Quantenelektronik (HQE), FB12, University of Siegen, 57068 Siegen, Germany (telephone: +49-271-740-2037, e-mail: francesco.voltolina@ieee.org), on leave from the University of Trieste and Sincrotrone Trieste.

Ralf H. Menk is with Sincrotrone Trieste, S.S.14, km 163.5, Basovizza, 34012 Trieste, Italy (telephone: +39-040-375-8201, e-mail: ralf.menk@elettra.trieste.it).

Sergio Carrato is with the Electrical Engineering Department (DEEI), University of Trieste, 34127 Trieste, Italy (telephone: +39-040-558-7147, e-mail: carrato@univ.trieste.it).

device, hence, this word size was considered convenient to represent the image pixels. The number of microstrips (1280) of the detector fixes the *column size*, while the *row size* is related to the maximum number of time slots allowed in a single measurement and its index is called `TIME FRAME` number.

The maximum number of TIME FRAMES (2048) is a compromise between the required *temporal resolution* in a measurement and the limited amount of RAM in which the DSP will have to store this image: 1280 x 2048 x 4 (32-bit) = 10 Mbyte that can easily fit in the 32 Mbyte Memory Bank of the DSP and will not be a limit for any practical measurement task. This is possible since we decided to make variable the number of `JAMEX Cycles` (having a fixed temporal length of 128 μs) per TIME FRAME and to store the relative values in another structure (`TFCycles`) that associates univocally a 32-bit integer to any possible TIME FRAME.

Hence, the longest measurement will have a duration limited to 2048 x $2^{32}$ x 128μs **= 2141 years** (cited just as a reference) while a measurement requiring the *highest temporal resolution* (limited by the JAMEX ASIC minimum integration time of 128μs) in *all* the TIME FRAMES must be performed on a phenomenon that is shorter than 2048 x 128μs = 262 ms, showing that all the available choices between these extremes can suit virtually any SAXS experiment.

### A. Real Time Acquisition and Processing

The Real Time data acquisition for a single measurement consists in getting 2048 TIME FRAMES, each consisting of *n* JAMEX CYCLES, making concurrent image processing activities, while providing intermediate results to monitor the progression, and building and storing the final image after the end of the measurement. The *main tasks* that have to be implemented in the Sundance System are:

- **JAMEX control** (generation of the main clock signals for the ASICs)
- **Image processing** (software normalization, correction and integration of the acquired data)
- **External devices control** (synchronizing all the activity with the EXTERNAL TRIGGER signals)

All the above main tasks involve specific design regarding the HOST software running on the PC Embedded and both the DSP software and the FPGA configuration of the Sundance System. The *Image Processing task* consists in building, from the Real Time acquisition of the JAMEX output signals, the matrix `Image[i,j]` (consisting of 1280 x 2048 words) in the DSP memory. This process is described from the following equations:

$$Image[i,j] = \sum_{k=1}^{TFC(i)} \{[JAMEXOUT(k+O(i),j) - Dark(j)] \cdot Norm(j)\}$$

where

$$O(i) = \sum_{n=1}^{i} TFC(n-1)$$

- *JAMEXOUT(p,q)* is the sampled value at the *p-th* JAMEX Cycle and relative to the *q-th* pixel (microstrip)
- *TFC(i)* is the *i-th* element of the vector TFCycles
- *Dark(j)* is the value defining the Dark Current subtraction for the *j-th* pixel of the detector
- *Norm(j)* is the value defining the Gain Normalization for the *j-th* pixel of the detector
- *O(i)* is a term introduced to simplify the notation, it represents the *JAMEX cycle offset* to the FIRST cycle of the *i-th* TIME FRAME

These equations show that the *i-th* TIME FRAME (represented as Image[i,*]) is the result computed on the DSP of the “software integration” performed adding on a per-pixel basis all the values sampled in the selected TIME FRAME. Hence, each resulting TIME FRAME is composed of 1280 JAMEX’s integrated samples, a value per pixel, on whom the DSP has applied a global dark current subtraction and normalization. These DSP computations will be repeated 2048 times, for each of the *i-th* TIME FRAMEs composing the matrix Image[*,*].

### B. Behavioural model of the SAXS detector

Having the aim to build a detector easy to familiarize with, was decided to model it in such a way that for a generic external user its operation would be no more complicated than controlling a simple state machine (see Fig. 2), which can exists in *three* fundamental states: **IDLE**, **READY** and **RUN**.

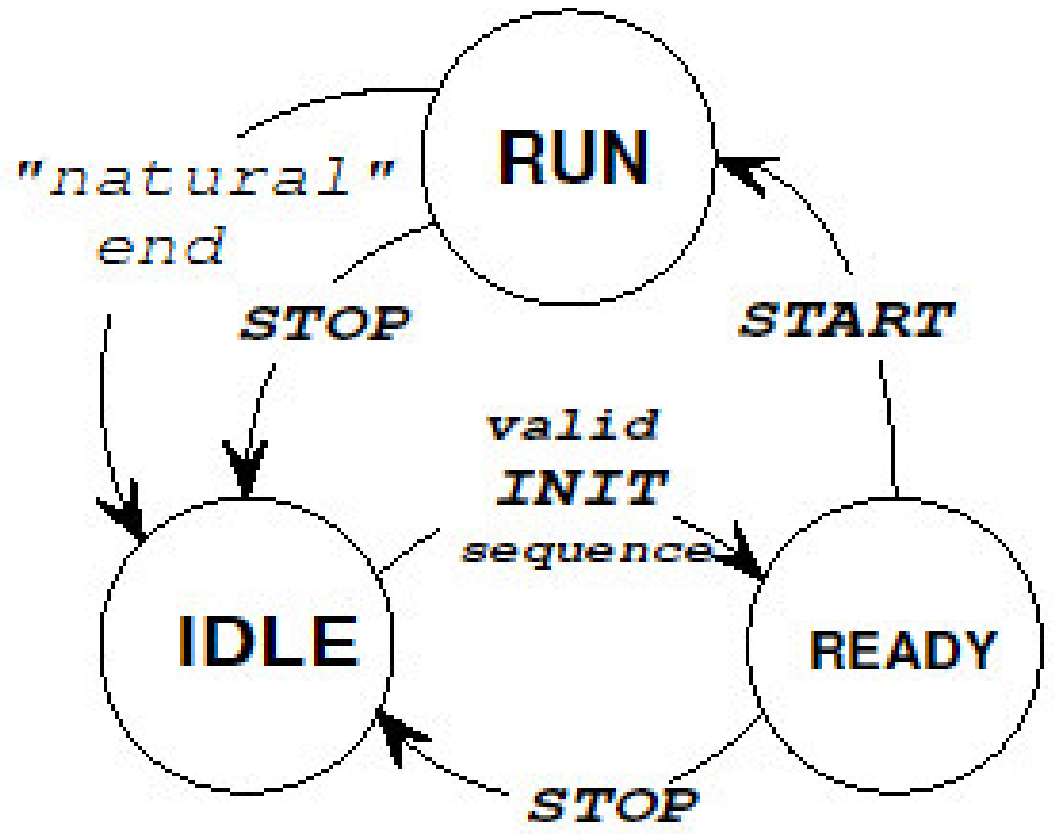

Fig. 2. The state machine representing the SAXS detector.

A state machine is an abstract but very effective way to describe the operation of basic up to very complex sequential devices. For our simple case a set of commands was created in order to control the transition between these states and, more importantly, the overall SAXS detector operation.

### C. User Commands

We will now introduce these commands showing a *typical sequence*, used to perform a complete measurement, with brief explanations aside and clearly indicating the state transitions:

**IDLE state**: state of the detector after power-up or when a measurement has completed

INIT PHASE: composed of many sequential sub-commands and leading, upon successful completion, to the READY state:

- `SET INTTIME` to set the 'time structure of our measurement'
- `SET TRIG OUT` defines when to generate 'trigger events'
- `SET TRIG IN` defines if to wait for any 'trigger events'
- `SET JAMEX CONFIG` to load the 'configuration bitstream' for the JAMEX
- `SET PGA GAINS` sets a specific gain for each of the 20 PGA
- `SET AUTO OFFSET` to initiate an automatic offset calibration of the 20 ADCs
- `SET COMP OFFSET` to specify an individual offset compensation for the 1280 channels
- `SET COMP NORM` to specify an individual normalization for each of the 1280 channels
- `SET REALTIME TX` to enable an automatic real-time transmission of the measured vectors
- `SET TESTMODE` to test the detector without X-rays via an integrated signal source

**READY state**: the detector is waiting and the `START` command can bring it to the RUN state

**RUN state**: the measurement and relative processing are performed. The following commands are allowed:

- `GET IMAGE` to see the actual image
- `GET VECTOR` to see the actual vector
- `GET ACTUAL LOOP` to see at which point of the measurement we are
- `GET ERR MESS` to see any eventual error message
- `STOP` that acts like BREAK when in RUN state

**IDLE state**: this state can be reached "naturally" when the programmed measurement task reached completion or via the `STOP` command that acts like an immediate BREAK, useful when, for example, the user discover to have set some wrong parameters and want to repeat the experiment without to wait.

All the USER COMMANDS created for this detector observe the rule that when a prefix is present, it states the membership to one of the *four categories* listed below:

- **ACTIVITY COMMANDS** including the two commands: START and STOP that are controlling the macroscopic activity of the detector
- **SET COMMANDS** all the commands of the INIT phase, they involve a transmission of data from the USER to the DETECTOR
- **GET COMMANDS** all these commands involve a transmission of data from the DETECTOR to the USER
- **WRITE COMMANDS** they exist for the only purpose of debugging eventual problems of the detector

## III. Hardware of the SAXS detector

The idea to choose a *low-power PC Embedded* as the host system for the Sundance acquisition system (see [4]), instead of a conventional PC, brought the design to the turning point that enabled the concept of `full-integrated SAXS detector` (see Fig. 3).

Under standard operative conditions, in addition to the main detector box, just an external POWER Supply (common to all the SAXS detector electronics, except the Gas-Gain section that will use a separated High-Voltage device), a standard 100baseT Ethernet cable and the optional external trigger cables will be required in the Experimental Hutch.

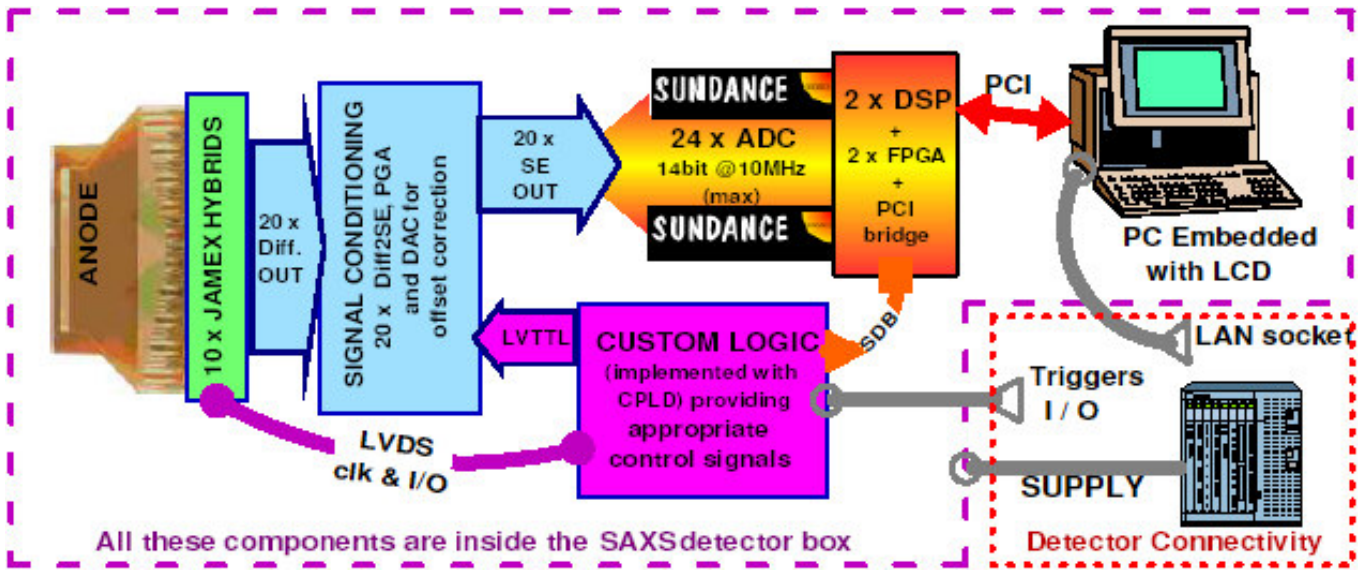

Fig. 3. Schematic view of the full-integrated SAXS detector.

This is a big quality step toward a detector that offers a good *ergonomic* to the users: it can be easily set in operation and positioned when needed, but also removed in short time and easily stored. The integrated PC embedded will run both a WEB and an FTP server making the detector to appear like a normal PC to which it is possible to "Login" in order to set the parameter for a new measurement or to download the results.

Two PCI cards will be contained inside the detector and carry the standard Sundance modules providing DSP and FPGA signal processing together with 24 analog inputs channels, implemented with single ADCs featuring 14 bit resolution at a sampling rate of 10 MHz. For the proposed configuration it was necessary to modify the JAMEX ASIC multiplexer in order to provide 64:1 multiplexing and one output port per chip: this was achieved modifying the *metallization mask* for the final production ASICs.

In this way, it is possible to connect the outputs of the JAMEX chips directly to 20 of the analog ADC inputs available while the spare 4 inputs will be used to monitor other important signals inside the detector box and for debugging purposes.

The J2S board (see Fig. 4) performs a differential to single ended conversion for the JAMEX outputs and an individually programmable offset shifting in order to use the full resolution of each ADC converter. Moreover, an analog first order low pass filtering is provided in order to reduce the overall system noise of the chain. A PGA (programmable gain amplifier) provides better flexibility for high dynamic range imaging and a built in limiting function helps to avoid ADC overflow recovery problems. Another important function of the J2S board is to provide the linear power supply, reference rails and the correct LVDS digital signals to control the operation of the 20 JAMEX ASICs, allowing functionalities like individual gain setting for each channel group and different operational modes. The cabling between the JAMEX Hybrids and the Sundance ADC cards take place through several connectors on this separate board.

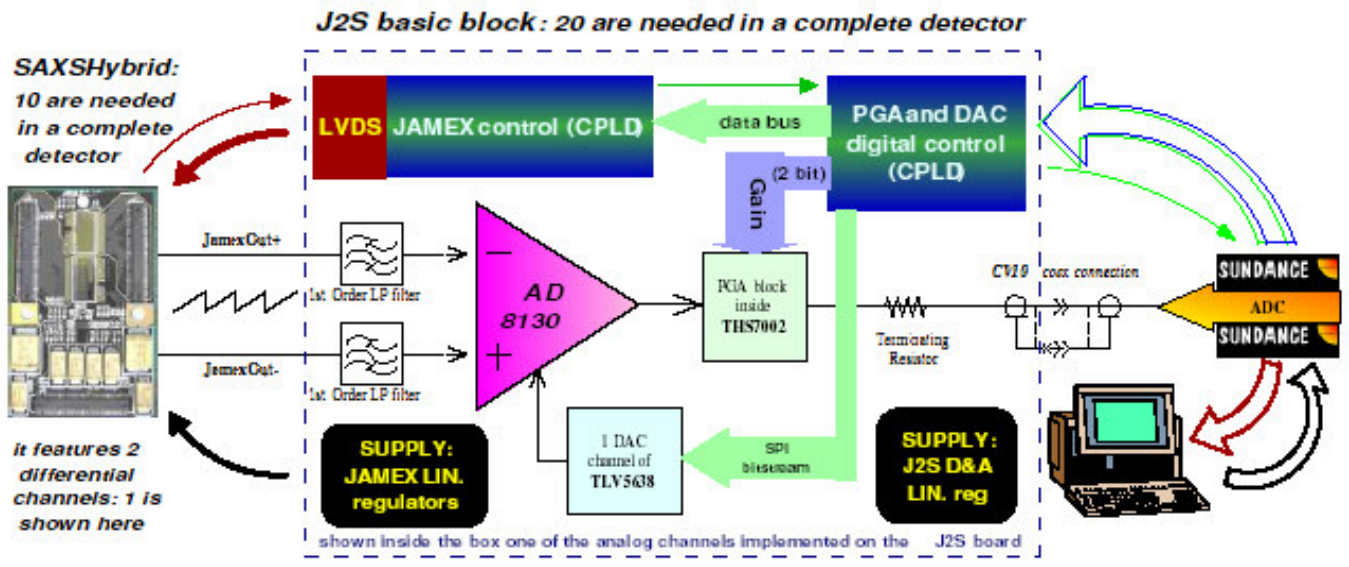

Fig. 4. Exemplification of the functionalities offered by the J2S board family.

Test signals generation capability is also embedded on the J2S board in order to give the possibility to test all the detector electronics without the need of an X-ray source. The decision to split the complex functionality of the original J2S board in many PCBs required some additional considerations in order to achieve a final configuration with a high degree of optimality.

Essentially, this configuration includes a PCB with the dimensions of a TIM-40 single Sundance module, called **J2S.SR**: it will be hosted on one Sundance carrier board and will be mainly responsible of buffering some clock signals for the ADCs, converting to LVDS the primary clocks for the JAMEX ASICs and offering some connectors aimed to an easier signal debug.

Another similar PCB will contain the optocouplers and isolated DC/DC converters necessary to offer the required *galvanic isolation* to the experimental trigger inputs and outputs: it will be placed in the other free place on the Sundance carrier board and will be called **J2S.OPTO**.

A third and bigger PCB will embed most of the functionalities of the original J2S board and will be produced in two versions called respectively **J2S.MINI** and **J2S.FULL**: the first one implement all the J2S digital control functionalities required by a full 1280 channels detector. On the other hand, it supports only 256 analog channels and its purpose is to proof the validity of the PCB design before to produce the final full version. This will be a 6-layers PCB implementing all the required analog channels and will lack of the debug connectors present in the J2S.MINI, in order to save space.

The last PCB in this series is the first that chronologically was produced: it is called **J2S.PIG**, meaning it is aimed to be mounted “piggyback” on the SAXS Hybrids.

### A. *The J2S.MINI board*

The J2S.MINI has been designed and built to *evaluate in a short time* the performance achievable from the components selected to build the final full-version called J2S.FULL.

It implements 4 analog J2S board channels (see Fig. 5): this circuit uses two key-components: the **AD8130**, a 270 MHz Differential Receiver Amplifier [5] from Analog Devices, and the **THS7002**, a programmable gain amplifier [6] from Texas Instruments. The JAMEX differential outputs are fed through a *first order RC low pass filter* to the high impedance inputs of the AD8130: the high CMRR of this device, approaching 70 dB at 10 MHz, allows the use of *unshielded flat cables* to transfer the JAMEX output signals without corruption by external noise sources or crosstalk.

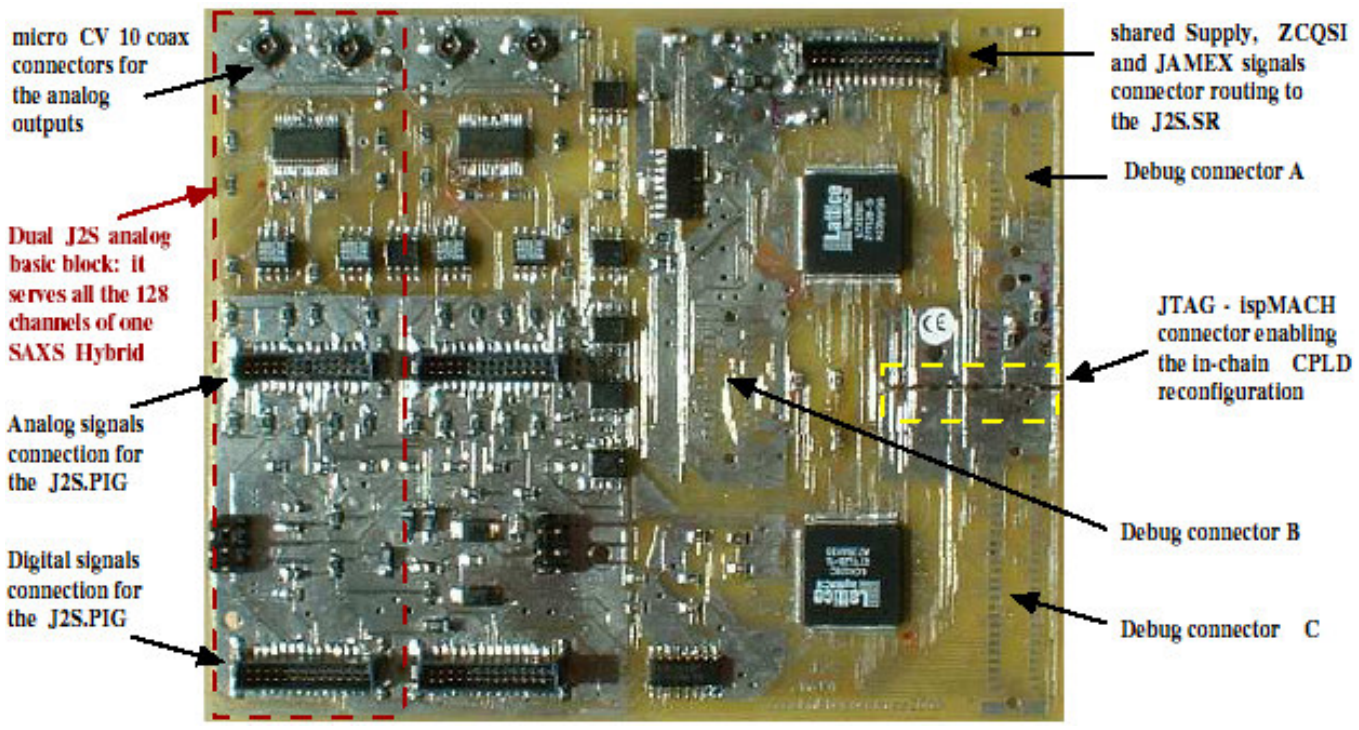

Fig. 5. Picture of the J2S.MINI front side with short explanations.

A definitive advantage of this device over a conventional operational amplifier is the ability to change the polarity of the gain merely by swapping the differential inputs: this characteristic is very useful in the J2S.MINI, where an unipolar positive DAC output is used to implement the adaptive offset subtraction feature without any additional component. The AD8130 output signal is directly connected to the 3-bit digitally controlled PGA that provides a **-22 dB** to **20 dB** attenuation / gain range with **6 dB** step resolution. In addition, the PGA provides separate high and low *output clamp protections* that are useful to prevent the output signal from swinging outside the input range of the ADCs present on the SMT356 modules.

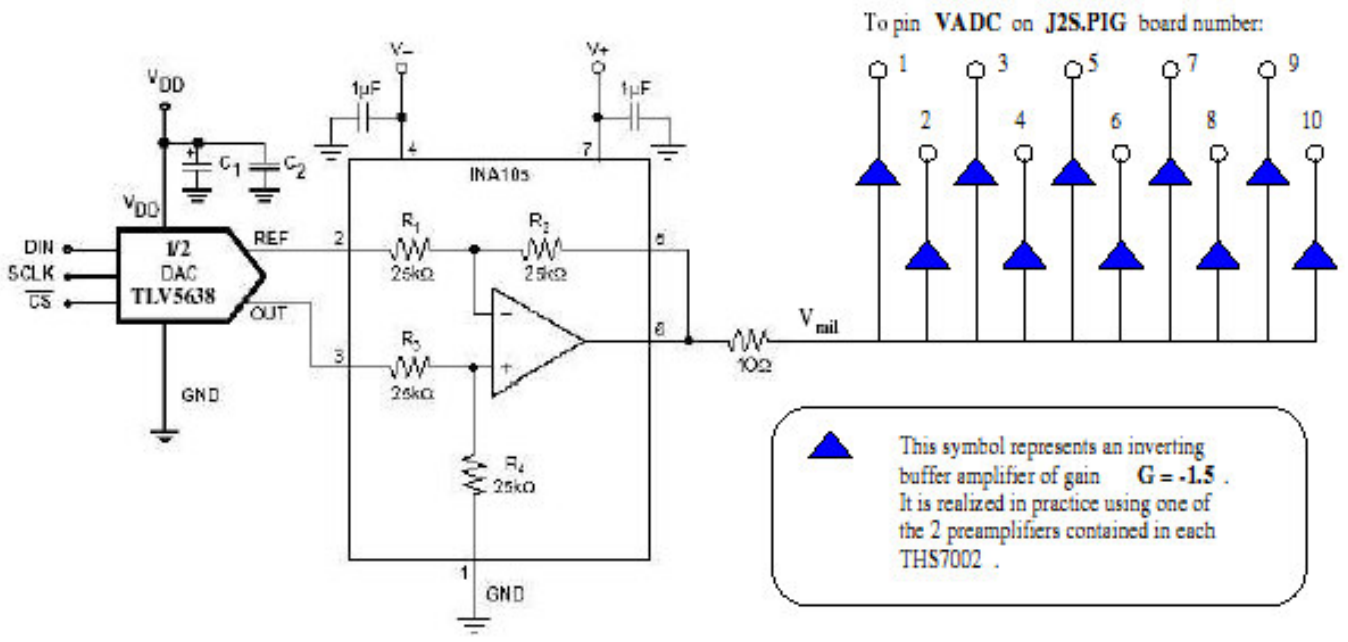

Fig. 6. The simple circuital configuration used to generate the required *variable reference voltage* for the JAMEX input stage negative supply: it feature a range of ±3 V.

This board implements all the additional logic to provide the 20 JAMEX ASICSs with the necessary signals for a correct operation (in connection with the Sundance system) during both the *Programming* and *Operating* mode.

This requires the implementation of a full featured *ZCQSI interface* and command set using a convenient CPLD: the **ispMACH 4128C** [7] from Lattice is here used in a compact 128 pin TQFP package. The relatively limited number of registers and I/O pins suggested the use of 2 of these devices in order to implement all the complex combinatorial logic, while allowing some more headroom for any modification, required in the future, that could be easily accommodated through the *In System Programming* offered from this device.

The chosen MACH CPLD does not embed any LVDS capability, but it is possible to fulfill those requirements using standard electronics components that perform the driver and receiver functions: the parts used in this design are the **SN65LVDM31** and **SN65LVDS32**, respectively a High-Speed Differential LINE DRIVER and a RECEIVER from Texas Instruments. In this way it is possible to control the JAMEX ASICs via any standard LVTTL digital logic device.

Not only the linear regulators for all the required analog and digital voltages are present on this board, but also some DACs, the low power SPI device TLV5638 from Texas Instruments [8], are employed in order to create the required voltage references (see Fig. 6) for an optimal operation of both the JAMEX ASICs and the integrated signal generator.

## IV. Summary and Conclusion

In many cases the electronics dictated the overall shape of the housing for the detector (custom built in aluminium), as in the case of the fixed dimensions of the Sundance acquisition system, but in general, and this is especially true if to consider the J2S board family, also the electronic design had been deeply influenced by the mechanical constraints.

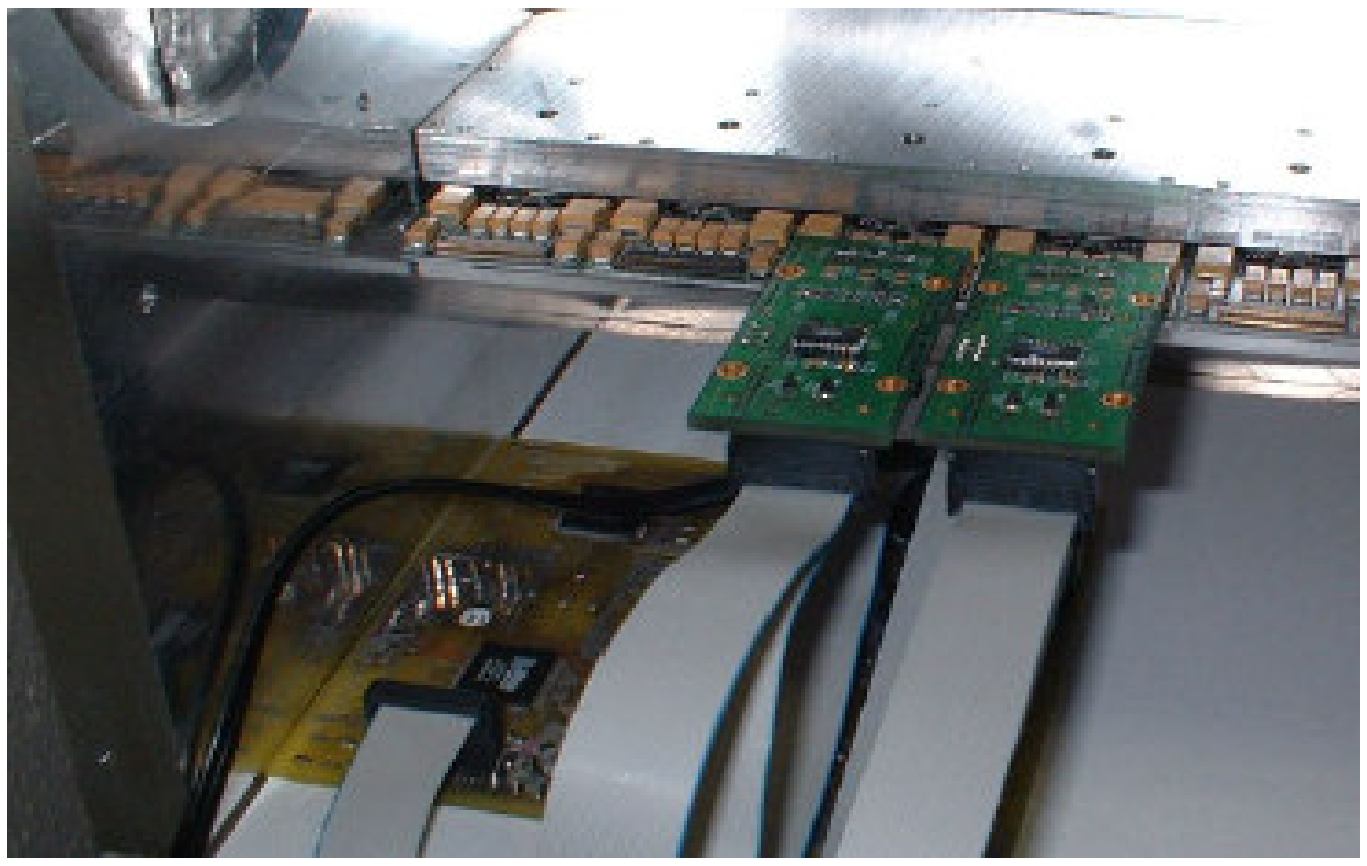

Fig. 7. The setup used to measure the JAMEX offset noise via the J2S boards connected to the Sundance acquisition system, not shown.

The main task, designing the circuit boards dedicated to interface the Sundance acquisition system to the JAMEX ASICs, reached a good level of completion with the production and test (see Fig. 7) of the definitive version of the J2S.PIG boards and of some well specified prototypes, called the J2S.MINI and J2S.SR boards, which can be developed in a definitive design with marginal changes.

On the side of the firmware and software required for the Sundance acquisition system, the question is still open, and a longer experimentation and some modifications are still needed to suit the final detector specifications.

## V. Acknowledgment

The authors would like to thank C.Fava from Sincrotrone Trieste for the design studies of the detector. The help of both the mechanical workshop and the electronics department of the University Siegen are gratefully acknowledged. Moreover, the authors would also like to thank all of those who are not mentioned by name but whose contributions are appreciated.

F.Voltolina is individually supported by the European Community – Research Infrastructure Action under the FP6 “Structuring the European Research Area” Programme (through the Integrated Infrastructure Initiative “Integrating Activity on Synchrotron and Free Electron Laser Science”).

The research presented in this article has been additionally supported by the European Community contract no. FMBICT980104.